\documentclass[journal=jpccck,manuscript=article]{achemso}

\usepackage[version=3]{mhchem} 
\usepackage[table,xcdraw]{xcolor}
\usepackage{graphicx}
\usepackage[table,xcdraw]{xcolor}

\author{Coulton Boucher}
\affiliation{Department of Materials Science and Engineering, McMaster University, 1280 Main Street West, Hamilton, Ontario L8S 4L8, Canada}
\author{Igor Zhitomirsky}
\affiliation{Department of Materials Science and Engineering, McMaster University, 1280 Main Street West, Hamilton, Ontario L8S 4L8, Canada}
\author{Oleg Rubel}
\email{rubelo@mcmaster.ca}
\affiliation{Department of Materials Science and Engineering, McMaster University, 1280 Main Street West, Hamilton, Ontario L8S 4L8, Canada}

\title[Application of murexide as a capping agent]
  {Application of murexide as a capping agent for fabrication of magnetite anodes for supercapacitors: experimental and first-principle studies}

\begin{document}

\begin{tocentry}

\begin{figure}[H]
  \includegraphics[height=4.3cm]{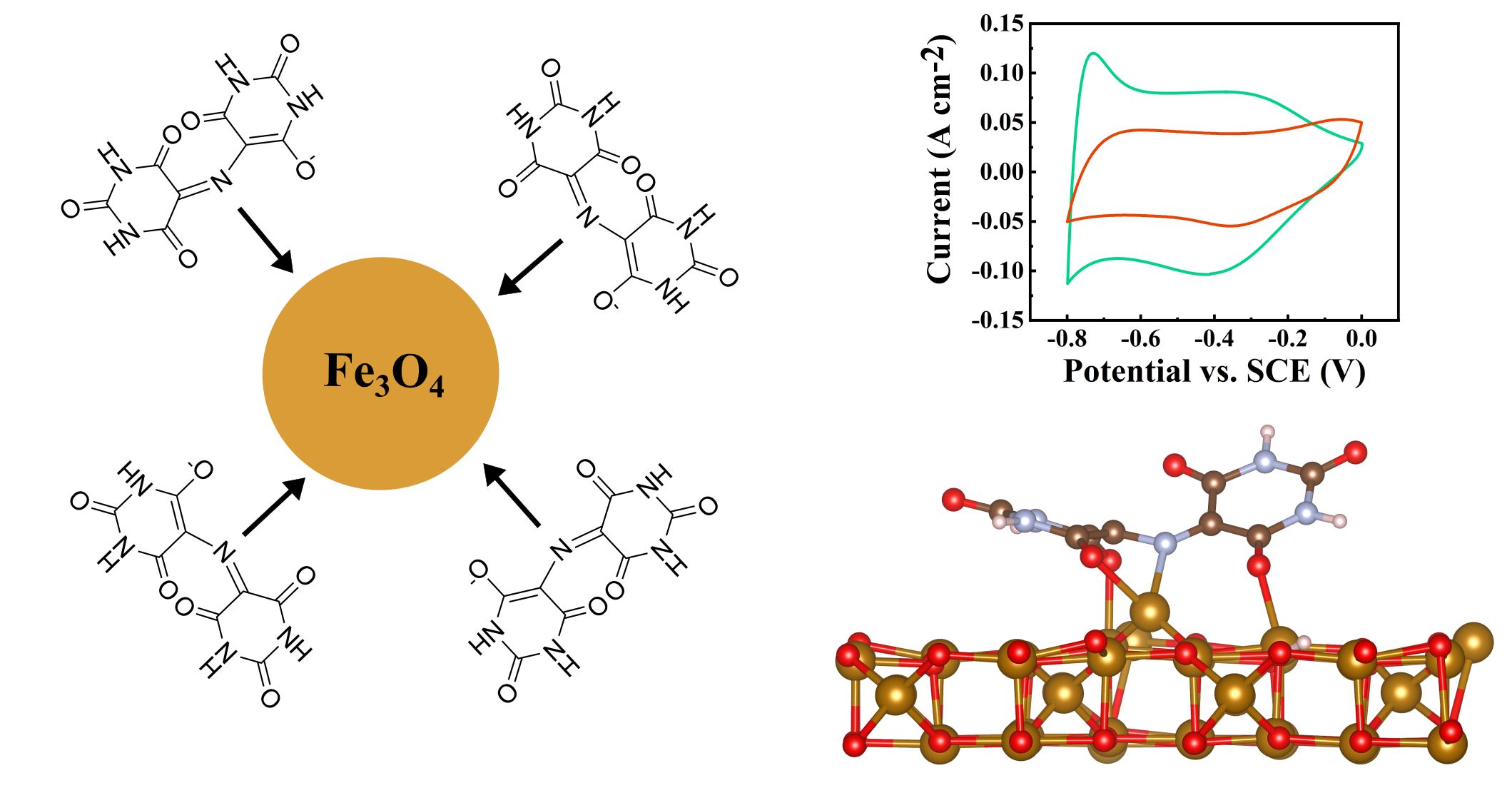}
\end{figure}

\end{tocentry}

\begin{abstract}
In this study, we investigate the effectiveness of murexide for surface modification of \ce{Fe3O4} nanoparticles to enhance the performance of multi-walled carbon nanotube-\ce{Fe3O4} supercapacitor anodes. Our experimental results demonstrate significant improvements in electrode performance when murexide is used as a capping or dispersing agent compared to the case with no additives. When murexide is used as a capping agent, we report a capacitance of $4.2$~F~cm$^{-2}$ from cyclic voltammetry analysis with good capacitance retention at high scan rate. From impedance measurements, we reveal a substantial decrease in the real part of impedance for samples prepared with murexide, indicating easier charge transfer at more negative electrode potentials, and reinforcing the role of murexide as a capping agent and charge transfer mediator. Density functional theory is used to investigate interactions between the murexide adsorbate and the \ce{Fe3O4}~(001) surface, with a specific emphasis on adsorption strength, charge transfer, and electronic properties. This theoretical investigation uncovers a strong adsorption enthalpy of $-4.5$~eV, and allows us to identify the nature of chemical bonds between murexide and the surface, with significant charge transfer taking place between the \ce{Fe3O4} surface and murexide adsorbate. The transfer of electrons from the \ce{Fe3O4} surface to murexide is recognized as a vital component of the adsorption process. By examining the bonding nature of murexide on \ce{Fe3O4}, this research study uncovers insights and proposes a novel bonding configuration of murexide that incorporates a combination of bridging and chelating bonding. 
\end{abstract}

\section{Introduction}
Supercapacitors have emerged as promising energy storage devices due to their high power density, fast charging and discharging rates, and long cycle life \cite{Olabi2022-nb}. Advanced electrode materials play a critical role in the electrochemical performance of supercapacitors, as they are responsible for storing and releasing charge during the charge/discharge cycles \cite{Forouzandeh2020-qm}. However, the electrochemical performance of these materials can be limited by factors such as low surface area, poor wettability, and high resistance at the electrode-electrolyte interface.

Surface modification has been identified as a key strategy to overcome these limitations and enhance the electrochemical performance of anode materials for supercapacitor applications \cite{Silva2018-uy,Zhao2015-vv}. Surface modification techniques such as surface roughening, the use of nanostructured materials, and the control of surface chemistry can increase the surface area, improve the wettability, and reduce the resistance at the electrode-electrolyte interface \cite{Pimsawat2019-ql,Yang2023-gg,Chen2011-et}. These modifications can lead to higher capacitance and energy density, improved rate capability, and cycling stability of supercapacitors.

By utilizing distinct materials for the anode and cathode, a supercapacitor device can optimize its operational voltage window by capitalizing on the unique potential ranges offered by each electrode \cite{Shao2018-tq}. Such an asymmetric device, containing \ce{Fe3O4} and \ce{MnO2} electrodes, can exhibit an expanded voltage window of 1.8~V in aqueous \ce{K2SO4} electrolyte and have a reported capacitance of 50~F~g$^{-1}$ at electrode mass density of 8.8~mg~cm$^{-2}$ \cite{Brousse2003-mm}. However, the negative electrodes display a significantly lower gravimetric capacitance than the positive electrodes, leading to a greater active mass in the negative electrode needed to match the capacitance of the positive electrode. This emphasizes the need to increase the specific capacitance of the negative electrodes, while ensuring excellent performance at high active mass. Designing electrodes with high active mass poses several challenges, including the poor charge transfer between \ce{Fe3O4} and conductive additives. Metal oxide nanoparticles and carbon nanotubes (CNTs) have high surface areas, making them prone to agglomeration, which further complicates the charge transfer process. Additionally, a decline in specific capacitance is observed as the active mass of capacitive material increases, primarily due to the weak electronic and ionic conductivities \cite{Gogotsi2011-xo} of metal oxide-based electrodes.

Recently, significant interest has been generated in application of chelating molecules for surface modification of materials \cite{Ata2014-rs}. The strong adsorption of such molecules on particle surface is an important factor for many applications. Molecules of different types are currently under investigation, including catecholates, gallates, salicylates, and other molecules from phosphonic acid and chromotropic acid families \cite{Ata2014-rs,Silva2018-uy}. Such molecules show strong bidentate or tridentate bonding to metal atoms on the particle surface, which is critical for their applications as capping agents for synthesis of nanoparticles, dispersing agents for colloidal processing and extractors for liquid-liquid extraction of nanoparticles\cite{Ata2014-rs,Silva2018-uy,Yang2022-bh}. Surface modification of metals with various catecholates such as Tiron and alizarin red, and molecules of other types such as chromotropic acid,  facilitated electron transfer, reduced electropolymerization potential,  and enabled electropolymerization of polypyrrole on non-noble substrates \cite{Tallman2002-kt,Shi2011-uk,Chen2013-aj,Ariyanayagamkumarappa2012-qd}. Of particular interest are applications of catecholates as photosensitizers for surface modification of semiconductors for photovoltaic applications \cite{Sakib2021-ya}. Catecholates were used as capping agents for synthesis of nanoparticles for supercapacitor electrodes with enhanced capacitive properties \cite{Yang2023-gg}. To enhance the capacitance of supercapacitor electrodes, a catecholate-type celestine blue molecule was employed as a capping agent during synthesis, a cationic dispersing agent for colloidal processing, and a charge transfer mediator \cite{Nawwar2020-jq}. 

Charge transfer mediators introduce a fast and reversible redox reaction that enhances the ionic conductivity and increases the pseudocapacitive capacity of the supercapacitor \cite{Yu2014-sg}. They can also store charges through valence changes and electron transfer between the mediators and electronic conductors, such as activated carbon, which provides additional charge capacity beyond that of electrostatic double layer capacitors \cite{Yin2011-vo,Akinwolemiwa2015-up}. Unlike pseudocapacitance, the redox mediator-induced capacitance is not dependent on the number of electrochemically active sites. The energy density of a redox mediator-based supercapacitor relies on the solubility of the redox mediators and their interaction with the electrodes. If highly soluble redox mediators are used, the volumetric energy density can be significantly improved as the dissolved mediator molecules or ions do not cause any significant volume changes in the entire system \cite{Tamirat2020-ch}. \citet{Roldan2011-ll} have shown that using hydroquinone as a redox mediator in an electrolyte supporting 1~M \ce{H2SO4} with a chemically activated carbon electrode can result in a two-fold increase in specific capacitance. This increase is due to the additional pseudocapacitive contribution from the Faradaic reactions of the hydroquinone/quinone system in the redox electrolyte. These studies have generated interest in the search for charged chelating molecules with redox properties for the development of advanced supercapacitor electrodes. 

Murexide is a versatile and widely used indicator in analytical chemistry. Murexide has been employed in complexometric investigations involving 3\emph{d} or 4\emph{f} ions, where the formation of a complex results in a modification of the solution's color \cite{Kashanian1988-ze, Shamsipur1989-ns, Shamsipur1989-pz,  Parham1992-ob}. It can form stable, colored complexes with a range of metal ions, including transition metals such as iron, cobalt, nickel, and copper \cite{Masoud2006-sx}. These complexes have distinctive hues that can be easily observed and quantified, making murexide an excellent tool for identifying and measuring metal ions in a sample. Furthermore, murexide can be employed in both aqueous and non-aqueous solvents, expanding its potential applications in a variety of fields \cite{Mohran2009-pr}.

Murexide has emerged as a promising capping and dispersing agent for the surface modification of cathode materials in supercapacitor applications \cite{Serwar-2017,Yang2022-bh}. Murexide has been used as a capping agent for \ce{Mn3O4} nanoparticles and as a co-dispersant for \ce{Mn3O4} and CNTs. The adsorption of murexide on \ce{Mn3O4} and CNTs facilitated electrostatic co-dispersion of \ce{Mn3O4} with CNTs with enhanced mixing to significantly increase the performance of the \ce{Mn3O4}-CNT electrodes \cite{Yang2022-bh}.

The redox properties of murexide have been investigated in various solvents and electrolytes, and the results have shown that it can exhibit multiple redox peaks with high redox potentials and good reversibility in aqueous electrolytes, indicating its potential for high energy density applications. In aqueous electrolytes, reversible cathodic and anodic peaks are well defined in a voltage window of 0 to $-1.5$~V vs. saturated calomel electrode (SCE) \cite{Mohran2009-pr}.

In this work, we are conducting the first density functional theory (DFT) study to investigate the adsorption mechanism and binding energy of murexide on a surface of \ce{Fe3O4}. The bonding mechanism of murexide on surfaces remains poorly understood, and therefore, we propose a novel bonding configuration that utilizes four atoms in the murexide molecule, forming a combination of bridging and chelating bonds. This proposed configuration contrasts with the tridentate \cite{Yang2022-bh,Masoud2006-sx,Atay2002ASS} bonding previously suggested in the literature and offers a promising explanation for the strong adsorption observed  experimentally. The charge density and Bader charge analyses are performed to understand the mechanism of molecular adsorption and magnitude of charge transfer during the adsorption process. The results of this study  provide insights into the underlying principles governing the interaction between murexide and \ce{Fe3O4} and will help to identify the optimal conditions for using murexide as an effective adsorbent in various applications. To corroborate the theoretical findings with experimental results, we have fabricated \ce{Fe3O4}-CNT composite electrodes using murexide as a capping and dispersing agent. This study aims to investigate, for the first time, the difference in electrode performance when the same molecule is used strictly as a dispersing agent, or as a capping agent. This combination of theoretical and experimental approaches provides a comprehensive understanding of the behavior of murexide as a dispersant agent, capping agent, and charge transfer mediator, and its potential use in energy storage applications.

\section{Results and discussion}

\subsection{Experimental results}

In this investigation murexide was investigated as both a dispersing and capping agent to facilitate the co-dispersion of \ce{Fe3O4} nanoparticles and CNTs to be used as the active material in supercapacitor anodes. Five techniques were conducted and tested where an electrode was fabricated from \ce{Fe3O4} and CNTs with no additives (NA), 5\% murexide added as a dispersing agent to synthesized \ce{Fe3O4} particles and CNTs in a solution of ethanol and dispersed in ethanol via probe ultrasonication (DE), 5\% murexide added as a dispersing agent to synthesized \ce{Fe3O4} particles and CNTs in a solution of water and dispersed water via probe ultrasonication (DW), 5\% murexide as a capping agent added during synthesis of \ce{Fe3O4} (C5), and 10\% murexide as a capping agent added during synthesis of \ce{Fe3O4} (C10). 

\begin{figure}[H]
\centering
  \includegraphics[width=\textwidth]{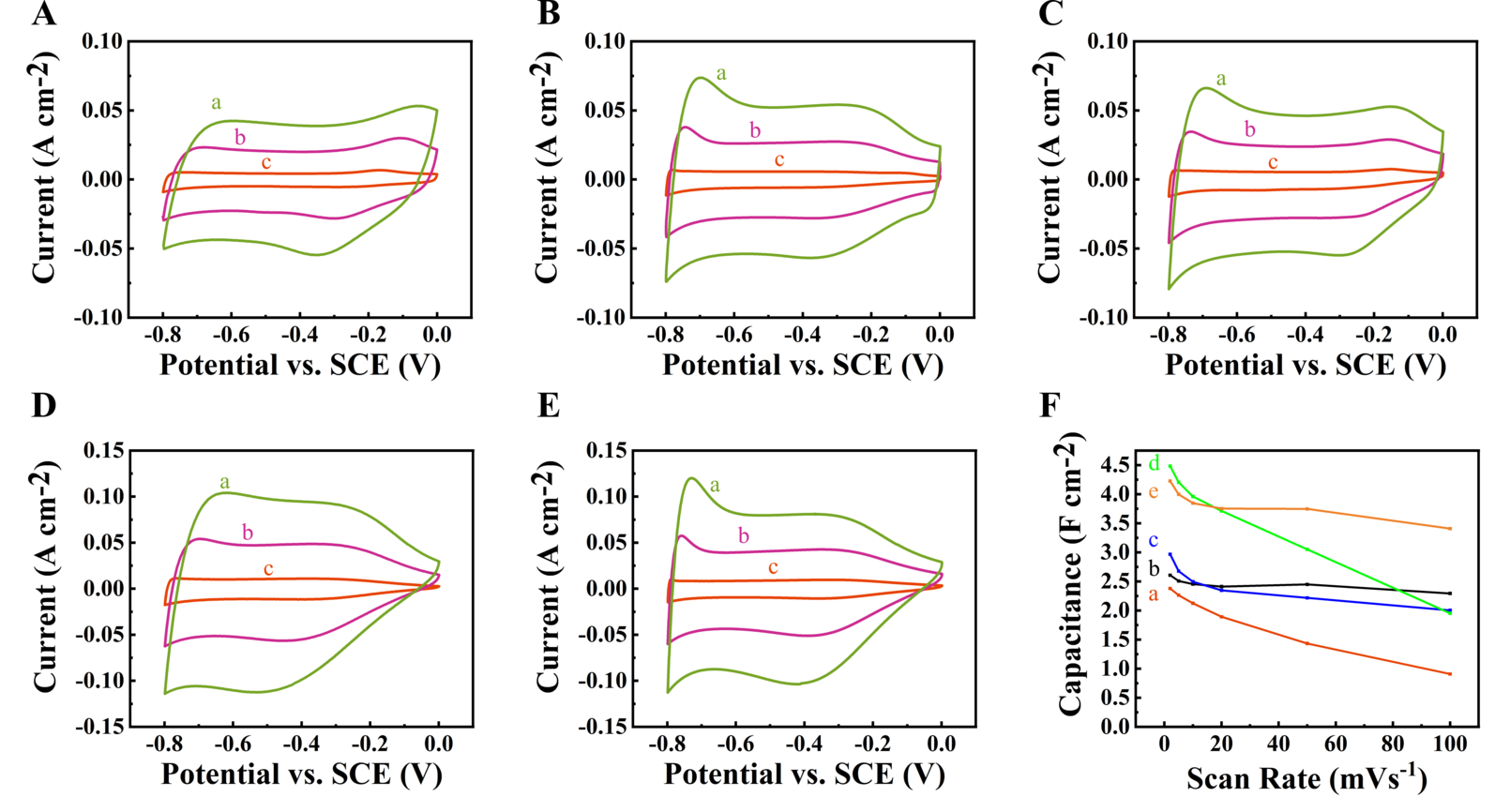}
  \caption{(A, B, C, D, E) CVs at scan rates of (a) 20, (b) 10, and (c) 2~mV~s$^{-1}$ for for (A) NA, (B) DE, (C) DW, (D) C5, and (E) C10. (F) Capacitance vs. scan rate for for (a) NA, (b) DE, (c) DW, (d) C5, and (e) C10.}
  \label{fgr:CV}
\end{figure}

The cyclic voltammetry (CV) curves in figure~\ref{fgr:CV} clearly show that the addition of murexide provides an increase in electrode performance compared to the case where no additive is used. In the cases where murexide is used as strictly a dispersing agent (DE and DW) we can see a slight increase in the peak capacitance of the electrodes from 2.4~F~cm$^{-2}$ for NA to 2.6~F~cm$^{-2}$ and 3.0~F~cm$^{-2}$ for DE and DW, respectively, at a scan rate of 2~mV~s$^{-1}$. We see capacitance retention at increasing scan rate improving greatly for DE and DW, where capacitance at 100~mV~s$^{-1}$ is 2.3~F~cm$^{-2}$ and 2.0~F~cm$^{-2}$, respectively, compared to 0.9~F~cm$^{-2}$ for NA. For C5 and C10, peak capacitance increases much more significantly, reaching a peak capacitance of 4.5~F~cm$^{-2}$ and 4.2~F~cm$^{-2}$, respectively, at 2~mV~s$^{-1}$ scan rate. The improved capacitance retention is not seen for C5 as capacitance fades to 1.9~F~cm$^{-2}$ at 100~mV~s$^{-1}$. When the concentration of murexide used as a capping agent increases from 5\% to 10\%, we see capacitance retention similar to DE and DW, where capacitance at 100~mV~s$^{-1}$ is reported at 3.4~F~cm$^{-2}$. 

We see a greater increase in the peak capacitance with murexide used as a capping agent compared to as a dispersing agent. We do, however, see better capacitance retention when murexide is used as a dispersing agent compared to a capping agent at a concentration of 5\%. 

When synthesis of \ce{Fe3O4} is conducted without additives, particles are able to agglomerate into larger particles after synthesis. Once murexide is introduced as a dispersing agent, these particles are no longer able to agglomerate further. This in combination with the adsorbed murexide playing the role of a charge transfer mediator results in a slight increase in peak capacitance and an improvement in the capacitance retention at higher scan rates. 

It is known that increasing the scan rate can result in a decrease in the measured capacitance \cite{Pell2001-nk}. This is because the scan rate affects the amount of time that the electrochemical reactions have to take place at the electrode-electrolyte interface \cite{Chowdhury2022-va,Gharbi2020-ab}. 
This results in the resistance at the electrode-electrolyte interface becoming a limiting factor in the charging and discharging processes at high scan rates \cite{Augustyn2014-qd,Chowdhury2022-va}. Murexide shows redox active peaks in the negative potential range, leading to a reversible 1 electron transfer process\cite{Mohran2009-pr} and allowing murexide to serve as a charge transfer mediator in a negative potential window. This charge transfer mediation allows for a decrease in resistance at the electrode-electrolyte interface. This means that electrons can more easily transfer during the surface redox reactions necessary for charge storage, which at high scan rates, need to happen very quickly in order to retain capacitance.

As a capping agent during synthesis, murexide is able to form complexes with the actively forming \ce{Fe3O4} surfaces. This prevents agglomeration as well as controls the growth of the \ce{Fe3O4} particles during formation. This results in a suspension containing smaller particles, which allows for the fabrication of an electrode with a higher active surface area for the necessary surface redox reaction to take place during cycling.

When the concentration of murexide as a capping agent is increased from 5\% to 10\%, there is more murexide to facilitate charge transfer from electrolyte to the active electrode material, effectively decreasing resistance at the electrode-electrolyte interface due to the redox properties of murexide. The redox properties of murexide will contribute indirectly to the capacitance of the electrode. This is evident in the samples where murexide is used as a capping agent (figure~\ref{fgr:CV} D and E) where we see an increase in the CV area specifically in the more negative potential range. The increase in CV area in only this range can be attributed to the influence of the adsorbed murexide molecule, which exhibits redox peaks in this range.

\begin{figure}[H]
\centering
  \includegraphics[width=\textwidth]{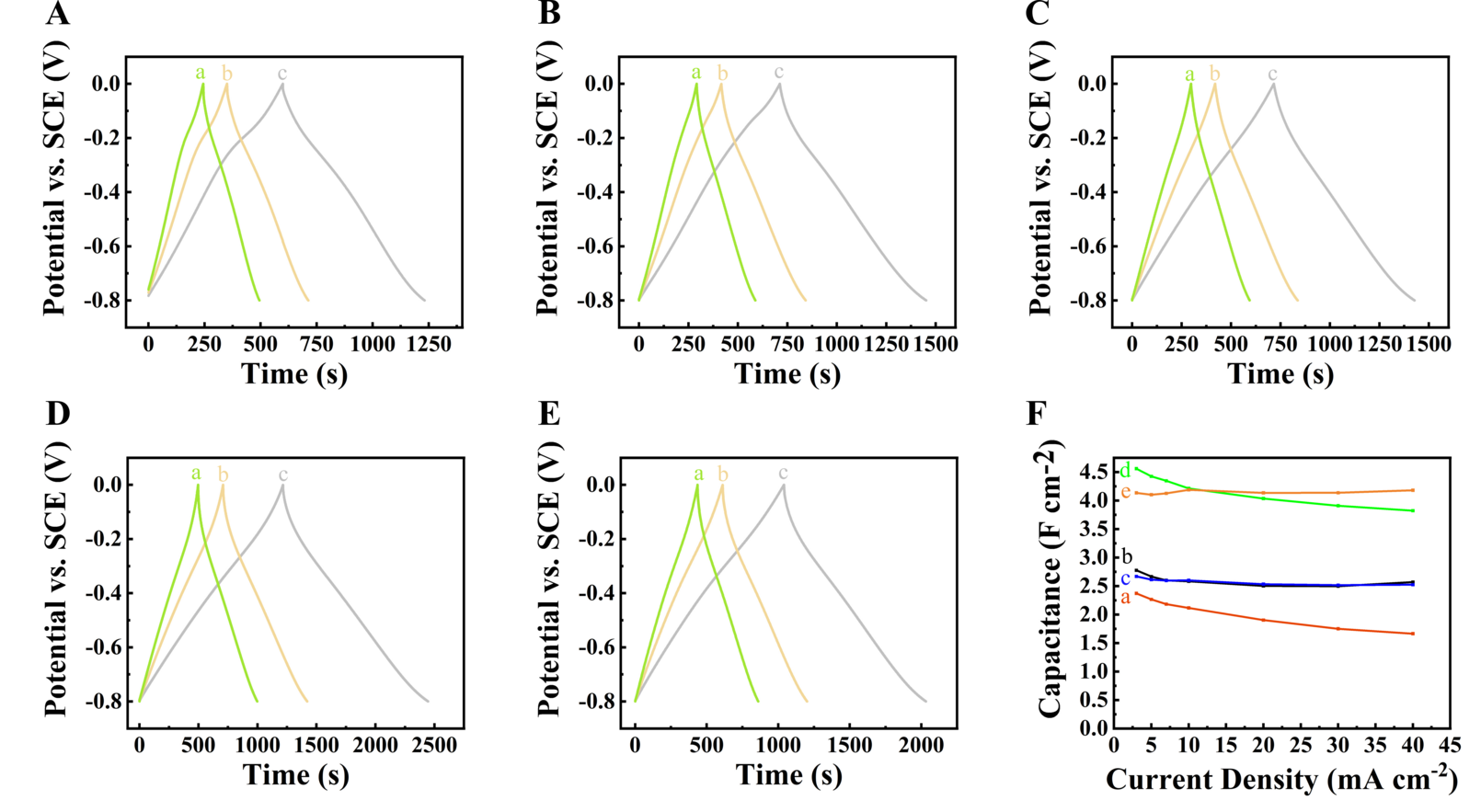}
  \caption{(A, B, C, D, E) charge-discharge curves at current densities of (a) 7, (b) 5, and (c) 3~mA~cm$^{-2}$ for (A) NA, (B) DE, (C) DW, (D) C5, and (E) C10. (F) Capacitance vs. current density profiles for (a) NA, (b) DE, (c) DW, (d) C5, and (e) C10.}
  \label{fgr:GCD}
\end{figure}

Figure~\ref{fgr:GCD} compares the galvanostatic charge-discharge (GCD) curves (panels~A-E) for different electrodes, with corresponding capacitance (panel~F) calculated from the discharge data at different current densities. GCD was performed in a potential window of $-0.8$~V to 0~V vs SCE. Nearly symmetrical and triangular charge-discharge curves are seen, confirming the pseudocapacitive behavior. Similar to the trends we see with the CV data, when electrodes showed only a slight increase in capacitance when murexide is used as a dispersing agent. For murexide as a dispersing agent in ethanol and water we see a peak capacitance of 2.8~F~cm$^{-2}$ and 2.7~F~cm$^{-2}$, respectively, at the current density of 3~mA~cm$^{-2}$ compared to a value of 2.4~F~cm$^{-2}$ for the case where no additive is used. When murexide is used as a capping agent, we again see a much more significant increase in the value of capacitance. For murexide as a capping agent with a concentration of 5\% and 10\% we see a peak capacitance of 4.6~F~cm$^{-2}$ and 4.1~F~cm$^{-2}$, respectively, at the current density of 3~mA~cm$^{-2}$. At current densities above 10~mA~cm$^{-2}$, we see that the capacitance of C10 exceeds that of C5.

\begin{figure}[H]
\centering
  \includegraphics[width=\textwidth]{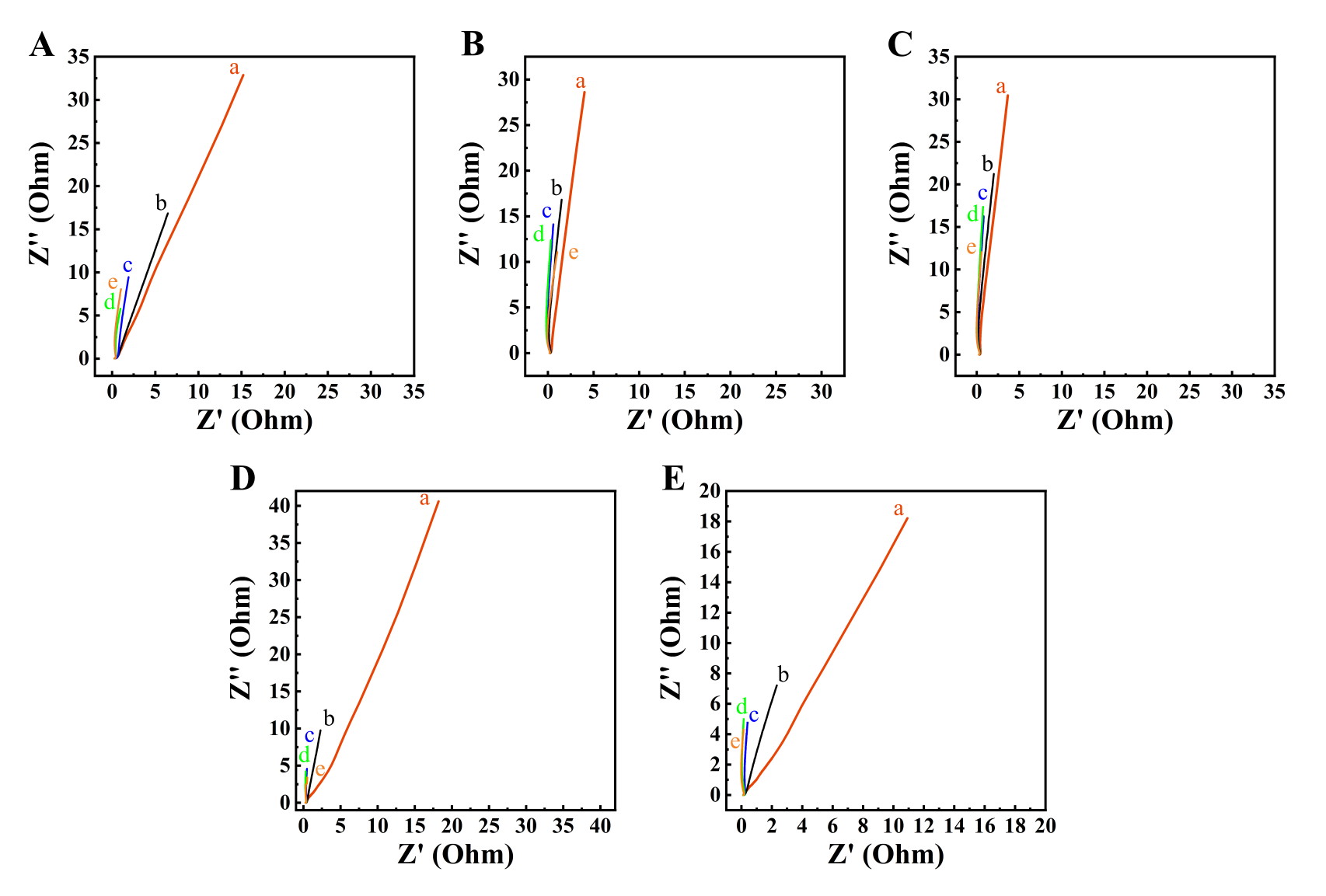}
  \caption{Nyquist plots for (A) NA, (B) DE, (C) DW, (D) C5, and (E) C10 at electrode potentials of (a) $0$, (b) $-0.2$, (c) $-0.4$, (d) $-0.6$, and (e) $-0.8$~V.}
  \label{fgr:Nyquist}
\end{figure}

The analysis of impedance data further provides evidence of the improved performance of the electrodes fabricated using murexide for surface modification of \ce{Fe3O4}. In figure~\ref{fgr:Nyquist}, we present the impedance at more negative potentials via Nyquist plots. We see a substantial decrease in the value of the real part of impedance for samples that were prepared with murexide, indicating that the presence of murexide results in easier charge transfer at more negative potentials. Although DE and DW samples exhibit the lowest real part of impedance at an electrode potential of $0$~V vs. SCE, we see that the C5 and C10 samples exhibit much lower impedance at more negative electrode potentials (below $-0.2$~V vs. SCE).  This decrease in impedance at lower electrode potentials is explained by the redox and electron mediation properties of murexide in this potential range.

\begin{figure}[H]
\centering
  \includegraphics[width=\textwidth]{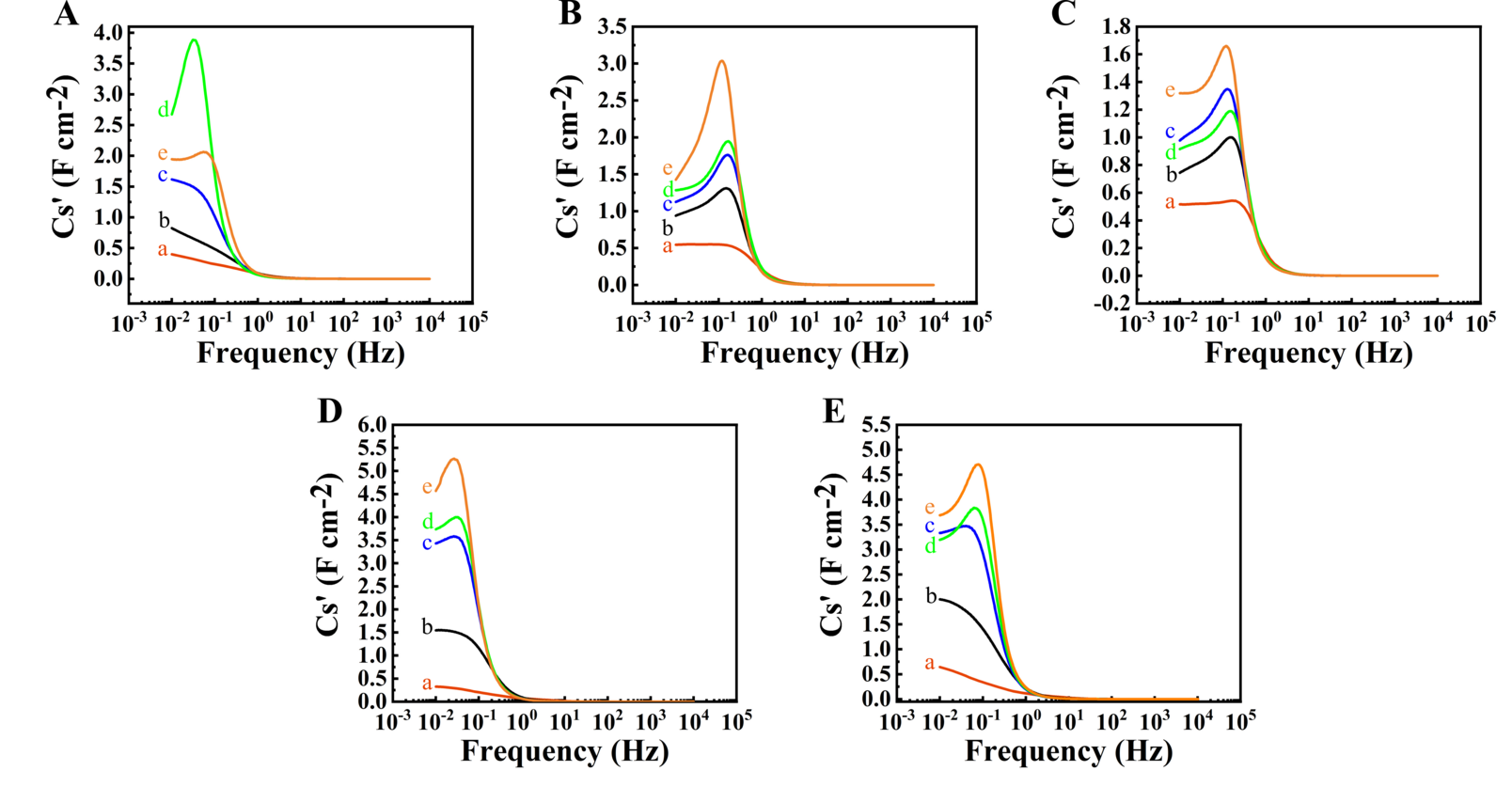}
  \caption{Real part of capacitance ($C'$) plots for (A) NA, (B) DE, (C) DW, (D) C5, and (E) C10 at electrode potentials of (a) 0, (b) $-0.2$, (c) $-0.4$, (d) $-0.6$, and (e) $-0.8$~V vs. SCE.}
  \label{fgr:Cre}
\end{figure}

Figure~\ref{fgr:Cre} shows the real part of capacitance for each of the five electrodes at $0$, $-0.2$, $-0.4$, $-0.6$, and $-0.8$~V vs. SCE. We see a significant improvement in the real part of capacitance ($C'$) for the C5 and C10 samples at electrode potentials below $-0.2$~V vs. SCE compared to other samples. This is consistent with what we see in the CV data, where the area of the CV curves drastically increases at lower electrode potentials when murexide is used as a capping agent. We see similar results with the dispersing agent samples (DE and DW) where the real part of capacitance is comparable to the case where no additive is used.

\begin{figure}
\centering
  \includegraphics[width=\textwidth]{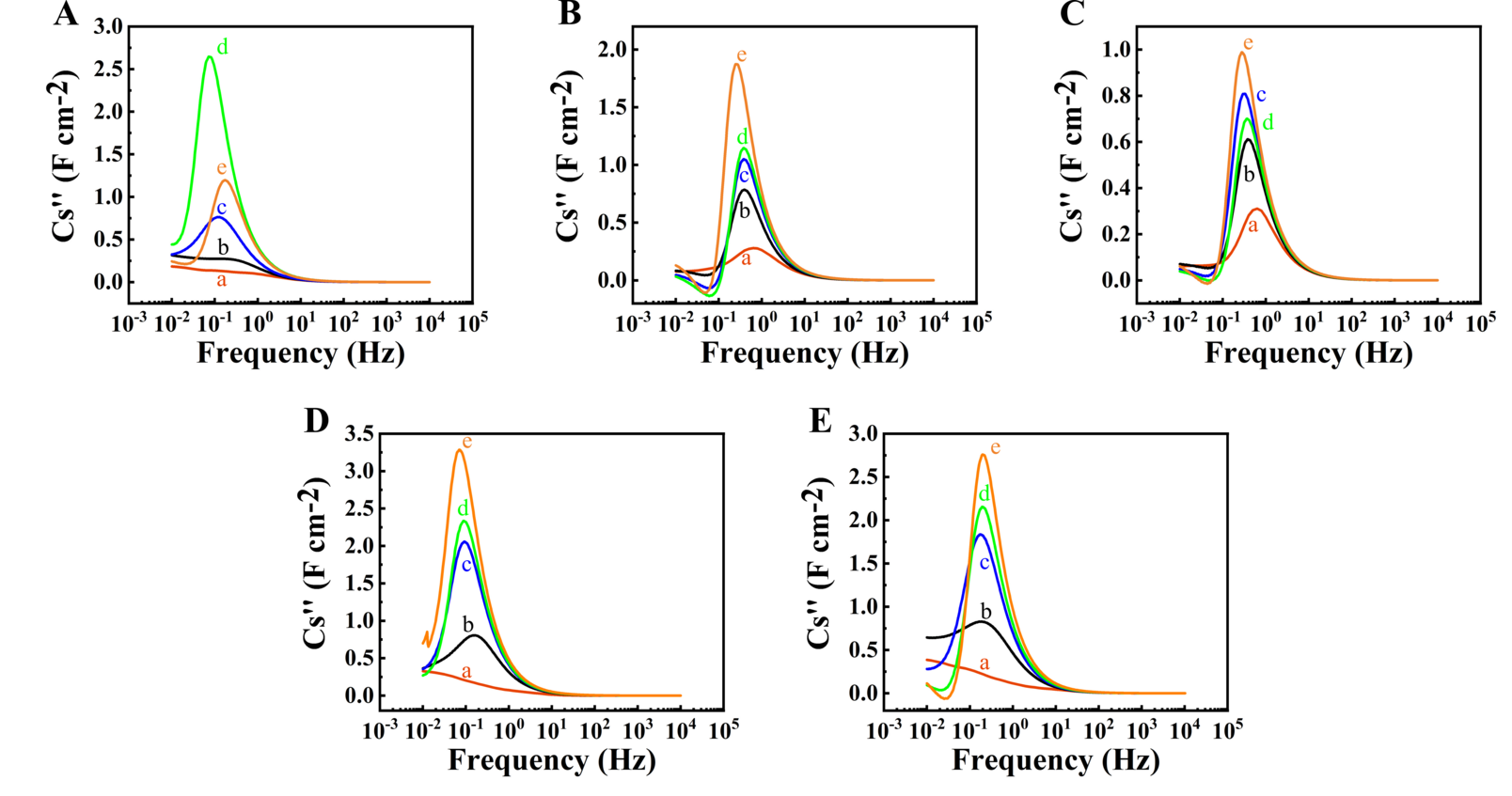}
  \caption{Imaginary part of capacitance ($C''$) plots for (A) NA, (B) DE, (C) DW, (D) C5, and (E) C10 at electrode potentials of (a) 0, (b) $-0.2$, (c) $-0.4$, (d) $-0.6$, and (e) $-0.8$~V vs. SCE.}
  \label{fgr:Cimg}
\end{figure}

From figure~\ref{fgr:Cimg} an increase of relaxation frequency from the imaginary part of impedance ($C''$) for the electrodes prepared with murexide as a dispersing agent (DE and DW) is observed. With the analysis of impedance at more negative electrode potentials, we see an increase in both the real part of capacitance (figure \ref{fgr:Cre}) and the relaxation frequency (figure~\ref{fgr:Cimg}) for samples prepared with murexide, indicating the indirect contribution to the overall capacitance of the electrode via the charge mediation properties of murexide. Additionally, the decrease observed in the real part of impedance (figure~\ref{fgr:Nyquist}) indicates relatively low resistance and the the large slope shown in the $Z''$ vs. $Z'$ curves shows good capacitive behaviour, especially at lower electrode potentials.

\subsection{DFT modelling of murexide adsorption on \ce{Fe3O4} (001) surface}

\begin{figure}
\centering
  \includegraphics[width=0.7\textwidth]{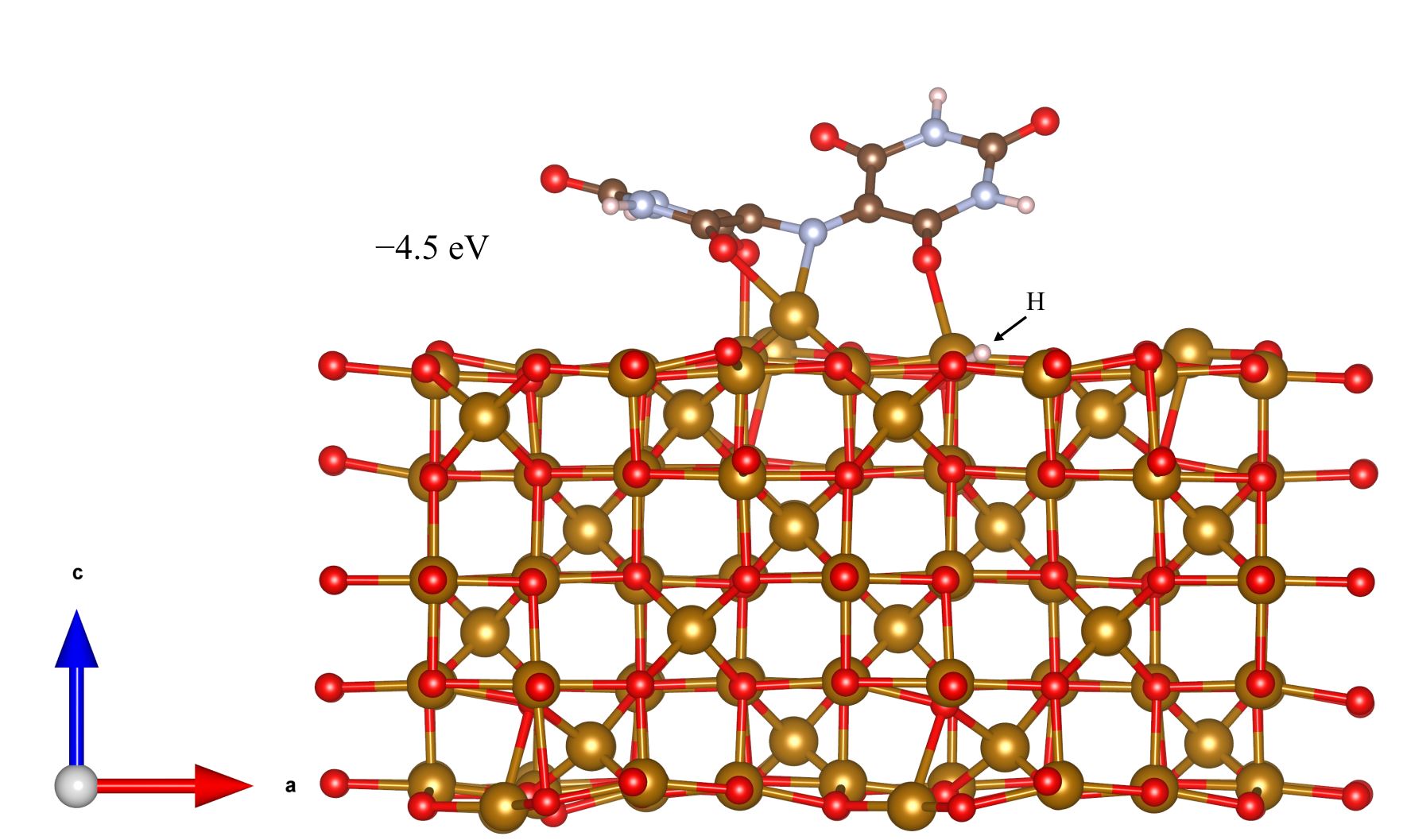}
  \caption{Adsorption of murexide on the (001) surface of \ce{Fe3O4}. The adsorption is accompanied by the surface adsorption of an \ce{H+} ion displaced from the OH group on the murexide molecule. The value of adsorption enthalpy is indicated.}
  \label{fgr:Hads}
\end{figure}

DFT modelling of the adsorption process of murexide on the surface of \ce{Fe3O4} is conducted. The structural model (in the lowest energy configuration) is illustrated in figure~\ref{fgr:Hads}. One hydrogen atom is cleaved from the murexide molecule and accommodated by \ce{Fe3O4} to form a surface OH group. Relaxation of bulk cubic \ce{Fe3O4} unit cell and subsequent construction of the (001) surface was performed following the procedure described elsewhere \cite{Boucher2023-ga}. The selected stoichiometric surface terminated with tetrahedrally coordinated Fe atoms, is in line with previous computational studies \cite{Bliem2014-ce,Gargallo-Caballero2016-yd} which have identified this (001) surface and its termination as the most energetically favourable.

Adsorption strength of murexide to the surface of \ce{Fe3O4} is evaluated by the calculation of an adsorption enthalpy ($H_{\text{ads}}$), which represents the difference between the total energy ($E_{\text{tot}}$) of the adsorbed and desorbed states
\begin{equation}
  H_{\text{ads}} = H_{\text{tot}}^{\text{ads}} - H_{\text{tot}}^{\text{des}}. \label{eqn:Hads}
\end{equation}
The desorbed state is modelled as the surface slab with the murexide molecule positioned 10~{\AA} above the surface in the vacuum layer. Our calculations yield an enthalpy of $H_{\text{ads}}=-4.5$~eV. This value indicates very strong adsorption, consistent with previous studies reporting adsorption enthalpies ranging from $-1.8$~eV for other organic molecules \cite{Boucher2023-ga} to $-5.5$~eV for adsorption of single adatoms \cite{Bliem2015-gx,Gargallo-Caballero2016-yd} on the \ce{Fe3O4} (001) surface. In comparison to other organic molecules, the adsorption enthalpy of murexide onto the surface of \ce{Fe3O4} exhibits a remarkable increase, attributed to the formation of four bonds with the surface. This significant enhancement in adsorption energy can be attributed to the unique bonding mechanism of murexide, which involves the simultaneous formation of both bridging and chelating bonds. The adsorption of murexide results in bonding to surface Fe atoms in a manner that maintains the octahedral or tetrahedral coordination, and distorts the surface to result in coordination of surface \ce{Fe} atoms closer to that of the bulk \ce{Fe} atoms in the \ce{Fe3O4}. 

\begin{figure}
\centering
  \includegraphics[width=\textwidth]{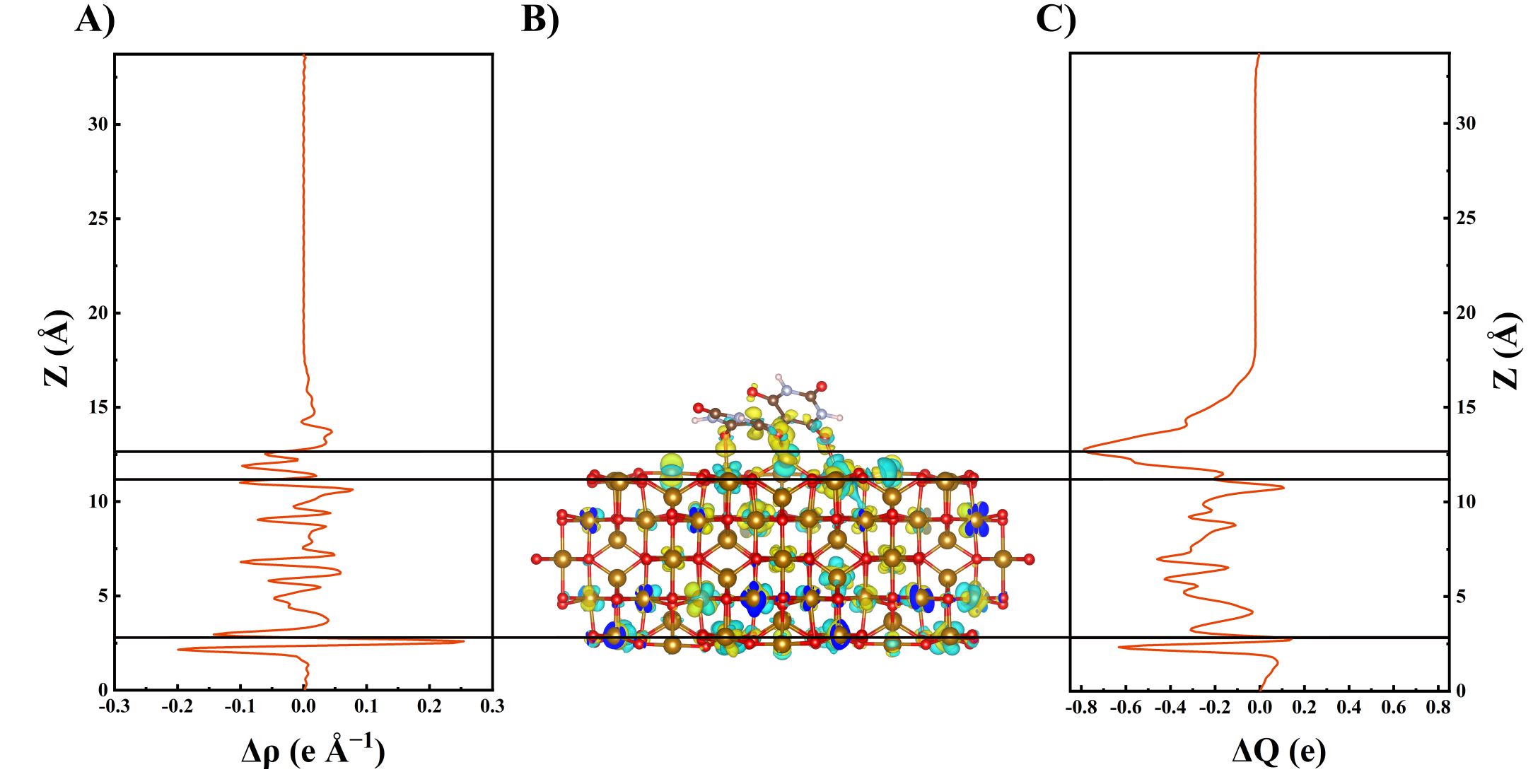}
  \caption{(A) Charge density difference planar average of murexide adsorbed on \ce{Fe3O4} surface, plotted along the $z$-axis, (B) schematic of adsorbed murexide on  \ce{Fe3O4} with three-dimensional isosurface of the charge density difference (rendered value of $\pm 0.0065$~e~{\AA}$^{-3}$), where yellow region represents area of electron accumulation and blue region represents area of electron depletion during the adsorption process, and (C) integral charge transfer $\Delta Q(z)$ with respect to position along the $z$-axis.}
  \label{fgr:CDD}
\end{figure}

In addition to the calculation of adsorption energies, the adsorption mechanism is also investigated using charge density difference plots and Bader charge analysis. Figure~\ref{fgr:CDD}(B) presents the three-dimensional charge density difference where the charge density of the separate surface ($\rho_{\text{surf}}$) and molecule, which includes the accommodated \ce{H}, ($\rho_{\text{mol}}$) is subtracted from the charge density of the adsorbed molecule on the \ce{Fe3O4} surface ($\rho_{\text{ads}}$)
\begin{equation}
  \Delta \rho = \rho_{\text{ads}} - \rho_{\text{surf}} - \rho_{\text{mol}}. \label{eqn:CDD}
\end{equation}
Figure~\ref{fgr:CDD}(A) plots the charge density planar average along the $z$-direction where the panel~(C) plots the amount of charge transferred up to $z$, and is given by
\begin{equation}
\Delta Q(z) = \int_{0}^{z} \Delta \rho (z)~dz.\label{eqn:Q}
\end{equation}

From figure~\ref{fgr:CDD}(A), we can see that the charge density around the \ce{Fe3O4} surface changes from positive to negative, indicating a transfer of negative charge from the \ce{Fe3O4} surface to the murexide molecule. From panel~(C) we see that the largest magnitude of charge transfer occurs at $z=13$~{\AA}, which is in the region where the bonds are formed between murexide and the \ce{Fe3O4} surface during adsorption. The total net number of electrons transferred between the \ce{Fe3O4} surface and adsorbed murexide is $0.79$~e, which implies an electron \textit{deficiency} of in the \ce{Fe3O4} slab due to the adsorbed murexide molecule.

\begin{figure}
\centering
  \includegraphics[width=0.7\textwidth]{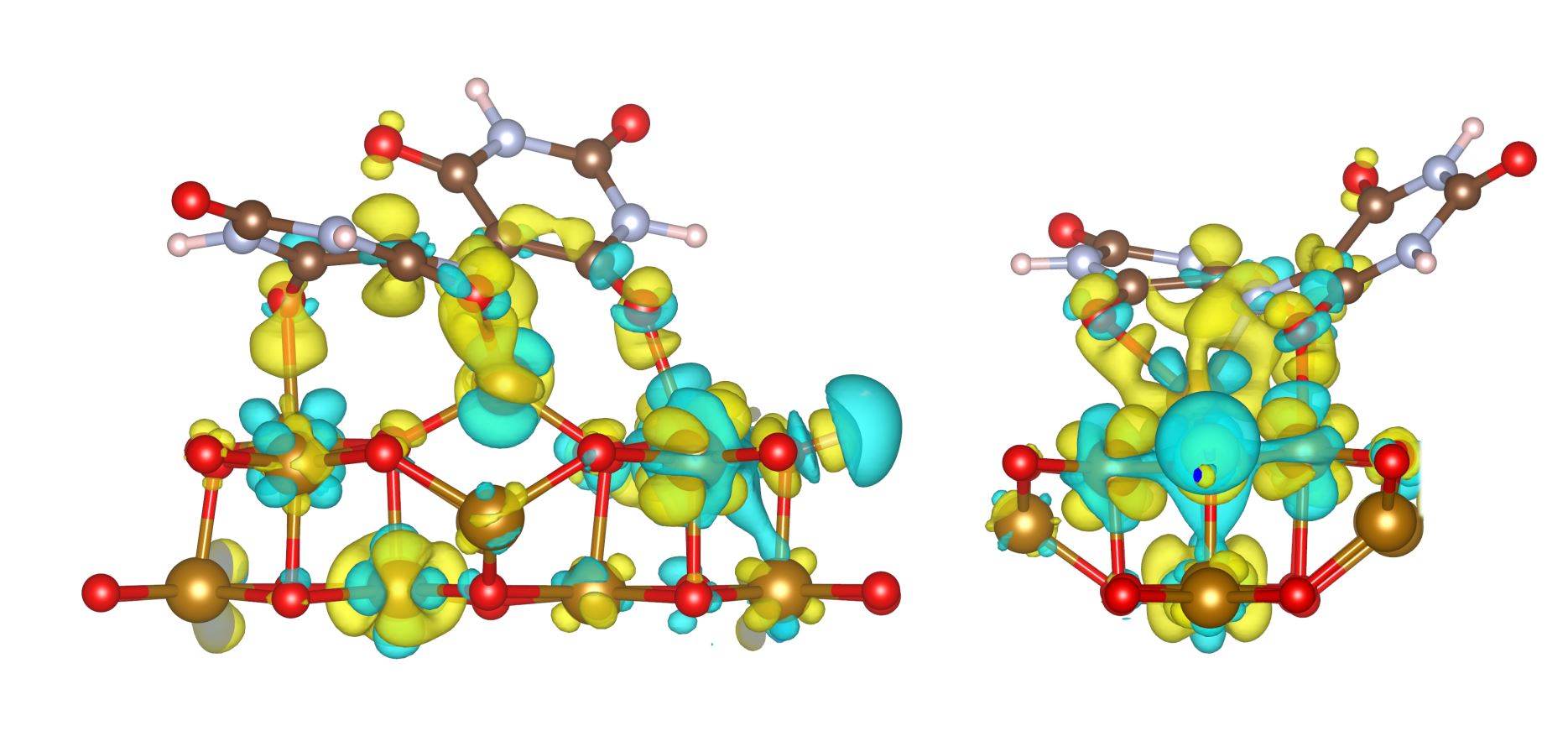}
  \caption{close-up view of three-dimensional isosurface of the charge density difference (rendered value of $\pm 0.0065$~e~{\AA}$^{-3}$) for atoms involved in murexide bonding to \ce{Fe3O4} surface, where yellow region represents area of electron accumulation and blue region represents area of electron depletion during the adsorption process.}
  \label{fgr:CDD_bonds}
\end{figure}

Since the charge density planar average is an average of the charge density at each point along the $z$-axis, we use Bader charge analysis \cite{Bader,HENKELMAN2006354} to further investigate the charge transfer between \ce{Fe3O4} and murexide during adsorption on an atom-by-atom basis. The Bader charges can be seen in table~\ref{tbl:Bader_murexide_side}, where we list the difference in Bader net atomic charges between the adsorbed molecule on \ce{Fe3O4} and the separate \ce{Fe3O4} surface and molecule. The bonds between murexide and \ce{Fe3O4} can be seen in figure~\ref{fgr:CDD_bonds} from the three-dimensional isosurface of charge density difference before and after adsorption. It is clear that electron accumulation occurs on the atoms involved in bonding that are part of the murexide molecule, where they are depleted from the surface Fe atoms, shown by the yellow and blue surfaces, respectively. This formation of a dipole results in the ionic nature of bonding of murexide onto the \ce{Fe3O4} surface.

By investigating the Bader charge difference between atoms in the adsorbed and desorbed states, we can gain insight into the charge transfer that occurs during the adsorption process, and which specific atoms play a role. From Bader charge analysis (table~\ref{tbl:Bader_murexide_side}), during the adsorption of murexide onto the \ce{Fe3O4} surface, we see a decrease in the Bader charge of $0.60$ for the murexide molecule with a subsequent increase of $0.62$ for the \ce{Fe3O4} surface. This indicates a depletion of electrons from the \ce{Fe3O4} surface with accumulation occurring in the murexide molecule. We can see this charge transfer takes place directly at the adsorption sites, where the bonds between the atoms in the murexide molecule form with the \ce{Fe} atoms in the \ce{Fe3O4} surface.

\begin{table}
\caption{Bader charge (in units of $|e|$) of selected atoms associated with murexide adsorption (see Fig.~\ref{fgr:Bond Labels} for labels).}
\label{tbl:Bader_murexide_side}
\begin{tabular}{lccc}
\hline
Location & Separate & Bonded & Difference \\
\hline
Surface                                         &                 &                &          \\
\quad Fe\textsubscript{1}                       &    1.58         &    1.73        &     0.15 \\
\quad Fe\textsubscript{2}                       &    1.33         &    1.66        &     0.33 \\
\quad Fe\textsubscript{3}                       &    1.58         &    1.70        &     0.12 \\
\quad Sum                                       &    ---          &    ---         &     0.60 \\
\hline
Murexide                                        &                 &                &            \\
\quad O\textsubscript{1}                        &     $-$1.00     &    $-$1.17     &    $-$0.17 \\
\quad O\textsubscript{2}                        &     $-$0.98     &    $-$1.10     &    $-$0.12 \\
\quad O\textsubscript{3}                        &     $-$1.04     &    $-$1.15     &    $-$0.11 \\
\quad N                                         &     $-$0.95     &    $-$1.17     &    $-$0.22 \\
\quad Sum                                       &     ---         &    ---         &    $-$0.62 \\
\hline
\end{tabular}
\\
\end{table}

\begin{figure}
\centering
  \includegraphics[width=0.8\textwidth]{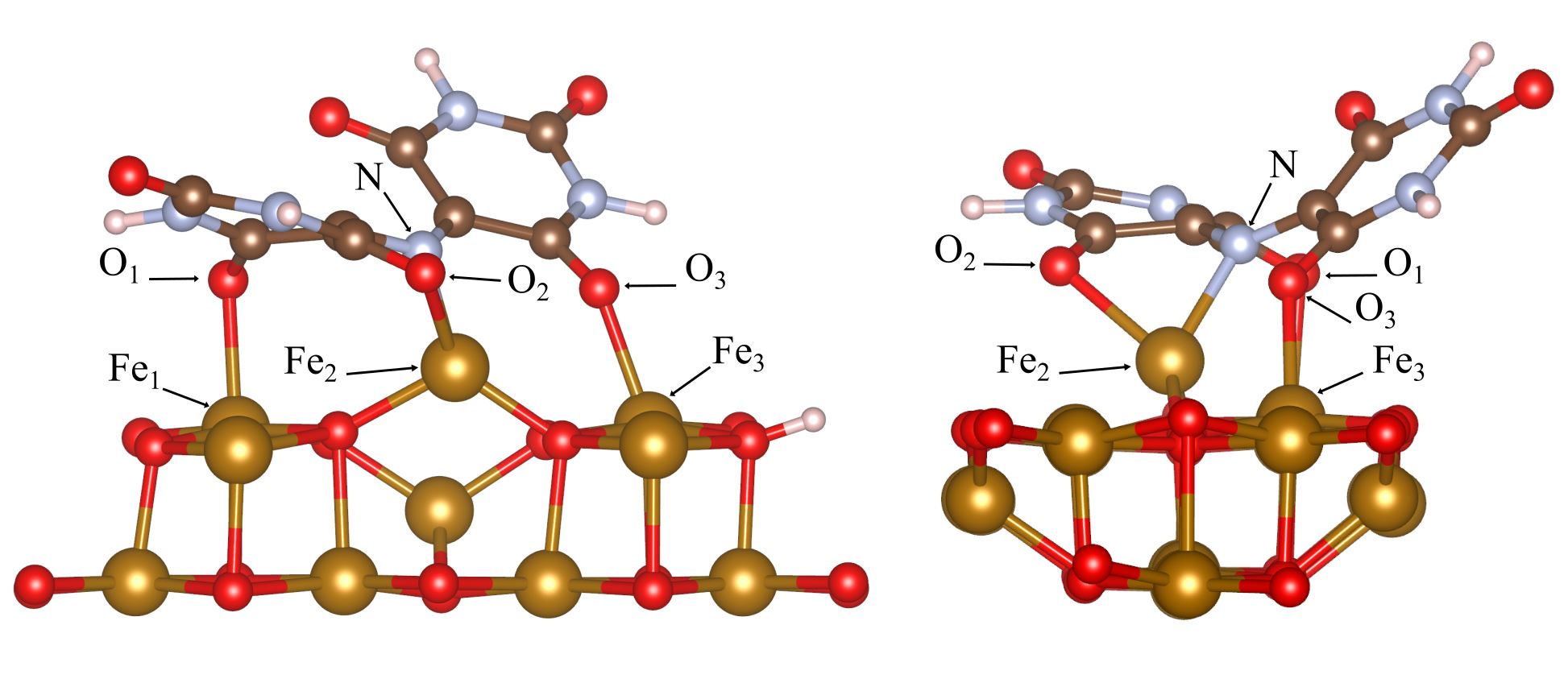}
  \caption{Close-up view of murexide adsorbed onto Fe$_3$O$_4$ surface with atoms labelled for Bader charge analysis.}
  \label{fgr:Bond Labels}
\end{figure}

From the charge density difference and Bader charge analysis, we can clearly see that the transfer of electrons from the \ce{Fe3O4} surface to murexide plays a critical role in the adsorption of murexide on \ce{Fe3O4}. Additionally, this pathway for electron transfer may provide insight into the charge transfer mediation that murexide provides in \ce{Fe3O4} supercapacitor anodes at negative electrode potentials.

\begin{figure}
\centering
  \includegraphics[width=\textwidth]{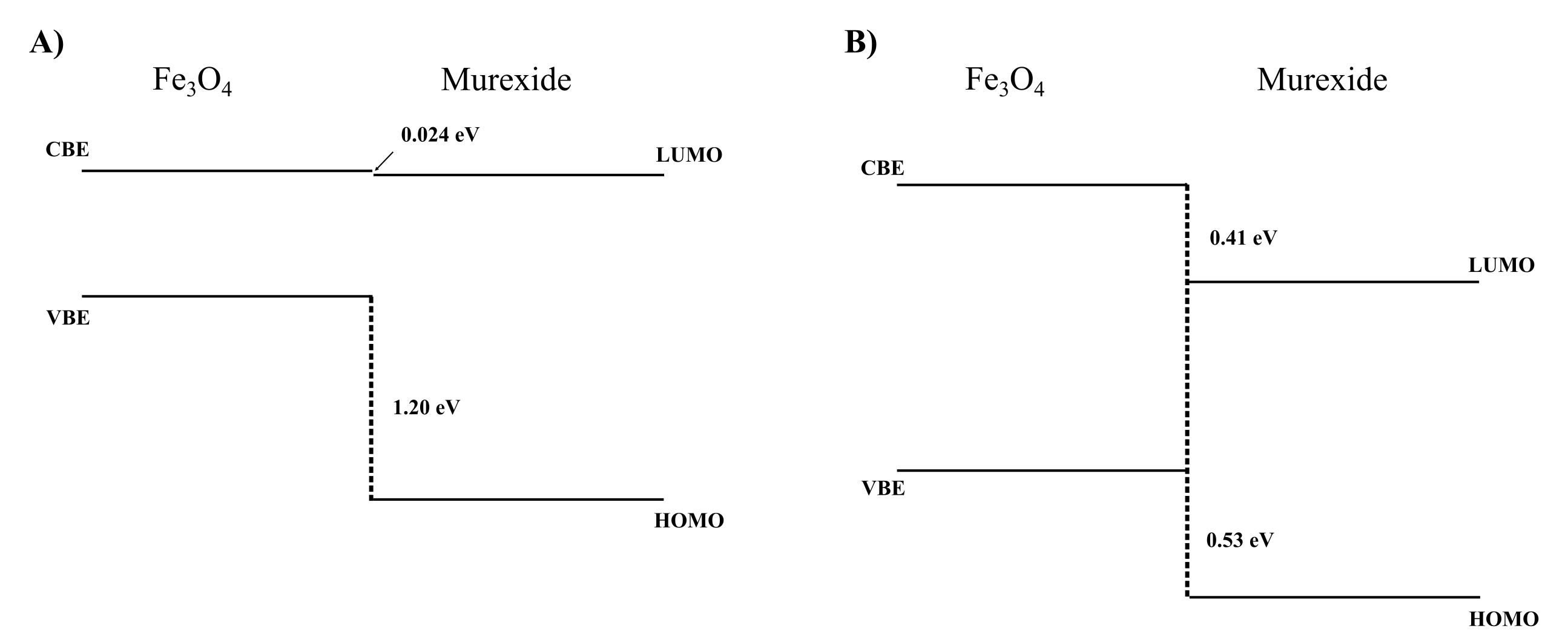}
  \caption{Schematic energy band alignment for (A) spin up and (B) spin down electronic states at the interface \ce{Fe3O4}/murexide.}
  \label{fgr:Band Alignment}
\end{figure}

From figure~\ref{fgr:Band Alignment}, we can see that, for both spins, the lowest unoccupied molecular orbital (LUMO), and the highest occupied molecular orbital (HOMO) of the murexide molecule are positioned lower in energy compared to the conduction band edge (CBE) and valence band edge (VBE) of the \ce{Fe3O4}, respectively. This energetic arrangement plays a pivotal role in electron transfer dynamics during the adsorption process. Specifically, the lower energy positioning of the murexide's LUMO facilitates the acceptance of electrons from the \ce{Fe3O4} conduction band, enabling efficient charge transfer from the metal oxide surface to the adsorbed molecule. This energetic alignment may also promotes enhanced electron transfer efficiency, with the possibility to increase the overall conductivity and performance of the system. 

\begin{figure}
\centering
  \includegraphics[width=\textwidth]{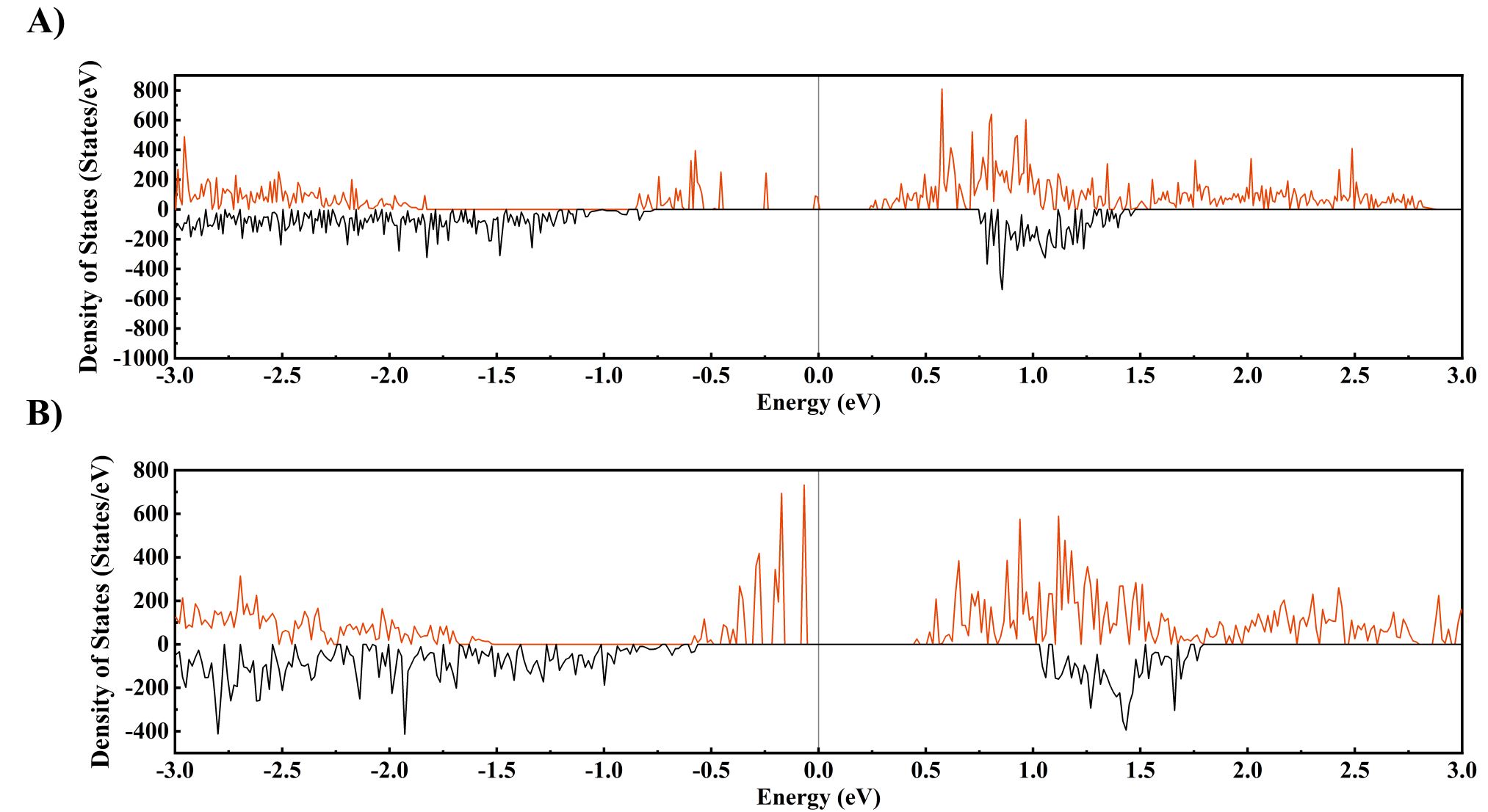}
  \caption{Density of states of the \ce{Fe3O4} surface before (A) and after (B) adsorption of murexide, where the red and black curves represent spin up and spin down, respectively. The Fermi energy is normalized to 0 eV.}
  \label{fgr:DOS_Surface}
\end{figure}

Further investigation of the density of states (DOS) reveals the formation of a band gap after the adsorption of murexide, while the surface is metallic (for one of the spin channels) in the non-adsorbed state (figure~\ref{fgr:DOS_Surface}). This band gap formation can be seen in figure~\ref{fgr:DOS_Surface}, where the density of states for the \ce{Fe3O4} surface before (A) and after (B) adsorption is shown. The formation of a band gap after murexide adsorption suggests a change in the electronic structure of the system. Adsorption-induced modifications, such as charge redistribution, can open up a band gap in the energy spectrum \cite{Rojas-Cuervo2023-db}. From charge-density and Bader charge analysis, the adsorption of murexide molecule on the \ce{Fe3O4} surface leads to a redistribution of charge. We see that the interaction between the murexide molecule and the surface atoms results in charge transfer, which can alter the occupancy of electronic states near the Fermi level. 

\section{Materials and Methods}
\subsection{Materials and Experimental Methods}
Iron (II) chloride tetrahydrate, iron (III) chloride hexahydrate, ammonium hydroxide, murexide (Ammonium 2,6-dioxo-5-[(2,4,6-trioxo-5-hexahydropyrimidinylidene)amino]-3H-pyrimidin-4-olate), poly(vinyl butyral-co-vinyl alcohol-co-vinyl acetate) (PVB, MilliporeSigma, Canada), multiwalled carbon nanotubes (MWCNT, ID 4~nm, OD 13~nm, length 1$-$2~$\mu$m, Bayer, Germany),  and nickel foam (porosity 95\%, thickness 1.6~mm, Vale, Canada), were used as starting materials.

Synthesis of \ce{Fe3O4} was performed by a chemical precipitation method \cite{Nawwar2020-jq,Nawwar2019-wg} using aqueous solutions of iron (II) chloride and iron (III) chloride. The molar ratio of \ce{FeCl2} to \ce{FeCl3} in the solutions was 1:2. 
For the synthesis of \ce{Fe3O4} in the presence of MWCNT, a 3~g~L$^{-1}$ MWCNT suspension was initially prepared. Preparation included ultrasonication of the MWCNT suspensions using a high energy Cole-Parmer (Canada) ultrasonic processor. Iron (II) chloride tetrahydrate and iron (III) chloride hexahydrate were added to the suspension, allowing for an \ce{Fe3O4}:MWCNT mass ratio of 4:1. The pH of the solutions was adjusted to pH=9 by ammonium hydroxide. Chemical precipitation was performed at 50~$^{\circ}$C at continuous stirring. Obtained suspensions were separated via centrifuge, washed, filtrated, and dried overnight in an oven at 60~$^{\circ}$C. For DE and DW samples, \ce{Fe3O4}-MWCNT suspension was redispersed in ethanol and water, respectively, via ultrasonication before being separated, washed, filtrated, and dried. For C5 and C10 samples, murexide was added to the Fe salt solution during synthesis in a weight percentage of 5\% and 10\%, respectively, then separated, washed, filtrated, and dried.

Electrodes were prepared by impregnation of Ni foam current collectors with slurries, containing \ce{Fe3O4}, MWCNT and PVB binder. The mass ratio of \ce{Fe3O4}:MWCNT:PVB was 80:20:3. The mass of the impregnated material after drying was 40~mg~cm$^{-2}$. The impregnated Ni foams were pressed to 30\% of their original thickness in order to improve electrical contact of the impregnated material and current collector. 

Electrochemical studies were performed in aqueous 0.5~M \ce{Na2SO4} electrolyte using Biologic VMP 300 potentiostat (BioLogic, France) for CV, electrochemical impedance spectroscopy (EIS), and GCD investigations. Testing was performed using a 3-electrode electrochemical cell containing a working electrode (impregnated Ni foam), counter-electrode (Pt mesh), and a reference electrode (SCE, saturated calomel electrode). Mass and area normalized capacitances were calculated from the corresponding CV and GCD data, as described by previous studies \cite{Shi2013-yj,Zhu2014-uy}. The capacitances calculated from the CV and GCD data represented integral capacitances measured in a potential window of $-0.8 \ldots 0$~V versus SCE. The capacitances calculated from the EIS data represented differential capacitances measured at a potential $0$~V, $-0.2$~V, $-0.4$~V, $-0.6$~V, and $-0.8$~V versus SCE, at voltage amplitude of 5~mV. CV results were obtained at 2, 5, 10, 20, 50, and 100~mV~s$^{-1}$ scan rates with EIS measurements performed afterwards. GCD results were obtained at 3, 5, 7, 10, 20, 30, and 40~mA~cm$^{-2}$ current densities.

\subsection{Computational}

The first-principles electronic structure calculations were performed in the framework of DFT \cite{Kohn1965-td} using Perdew–Burke–Ernzerhof (PBE) generalized gradient approximation \cite{Perdew1996-cl} for the exchange correlation functional, augmented by the DFT-D3 correction with Becke–Johnson damping \cite{Grimme2011-rk,Grimme2010-vp} to capture van~der~Waals interactions. The Vienna \textit{ab initio} simulation program (VASP) (version 5.4.4, University of Vienna, Vienna, Austria) \cite{Kresse1993-oq,Kresse1996-fe,Kresse1996-mj} and projector augmented-wave potentials \cite{Kresse1999-ub} were used, where the $p$~semi-core states were treated as valence states for Fe potentials in all calculations. Standard potentials were used for all other elements. The cut-off energy for a plane wave expansion of 400~eV was used for adsorption calculations. We included on-site Coulomb interaction to treat the highly correlated Fe 3$d$-electrons in the framework of \citet{Dudarev1998-dq} using an effective Hubbard energy of $U=3.7$~eV \cite{Bliem2015-gx}. Collinear spin-polarized calculations were performed for all structures. Magnetic moments were initialized with opposing spin orientations of magnitude 4.0~$\mu_B$ for tetrahedrally and octahedrally coordinated Fe atoms \cite{Chiter2016-qy}. Only forces were relaxed for surface models of \ce{Fe3O4} with an additional constraint of atomic position for the middle three layers of the \ce{Fe3O4} surface structure, to maintain bulk atomic positions. The structure was considered as optimized when the magnitude of Hellmann–Feynman forces acting on atoms dropped below 50~meV~{\AA}$^{-1}$ and components of the stress tensor did not exceed 1~kbar. The ground state energy was calculated using first order Methfessel–Paxton smearing with a width of 0.02~eV. A blocked-Davidson algorithm with high precision is used during the relaxation of the bulk and surface \ce{Fe3O4} structures. The Brillouin zone was sampled with a $\Gamma$-centered k-mesh generated automatically with a linear density of 30 divisions per 1~{\AA}$^{-1}$ of the reciprocal space.

The electronic and charge properties of the murexide adsorbate and \ce{Fe3O4} system was investigated by analyzing the Bader charge \cite{ HENKELMAN2006354} and DOS analysis. In Bader analysis, the electron charge distribution from the DFT calculation was partitioned and assigned to individual atoms. The differences in the partitioned charge before and after adsorption indicate charge transfer between the surface and adsorbate. DOS analysis examines chemical bonding interactions by showing the changes in the occupation of the electron energy levels associated with adsorption. Charge density planar average and DOS plots were obtained from data using VASPKIT \cite{ Wang2019-mv}. 

All structure files and VASP input files used in this work can be found in the Zenodo file repository \cite{Zenodo_10.5281/zenodo.8183854}. Structure files can be visualized in VESTA \cite{Momma2011-pf}.

\section{Conclusions}

In conclusion, the experimental results presented in this study provide evidence for the effectiveness of murexide as a capping agent to enhance the performance of MWCNT-\ce{Fe3O4} supercapacitor anodes. The addition of murexide, whether as a dispersing agent or a capping agent, resulted in significant improvements in electrode performance compared to the case without any additive. When used as a dispersing agent, 5\% murexide in ethanol and water led to slight increases in peak capacitance and remarkable improvements in capacitance retention at higher scan rates. On the other hand, murexide as a capping agent during synthesis resulted in a more substantial, 1.9-fold increase in peak capacitance, reaching values as high as 4.6~F~cm$^{-2}$ in the case of the C5 sample, and maintaining the improved capacitance retention when murexide concentration is increased.

The analysis of impedance data further supported the enhanced performance of murexide-modified electrodes, showing easier charge transfer and improved capacitance at negative electrode potentials. GCD data confirmed the pseudocapacitive behavior of the electrodes and demonstrated the superior charge storage capacity of murexide-modified electrodes.

The atomistic modelling of the adsorption process of murexide on the \ce{Fe3O4} surface provided valuable insights into the adsorption strength, bonding characteristics, charge transfer, and electronic properties. The strong adsorption of murexide on the \ce{Fe3O4} surface is indicated by an adsorption enthalpy of $-4.5$~eV. Additionally, it is evident that the coordination of \ce{Fe} atoms in bonding significantly influences the interaction between atoms. Specifically, \ce{O} and \ce{N} from murexide forms bonds with surface \ce{Fe} atoms in a manner which restores their respective bulk tetrahedral or octahedral coordination. Further analysis using charge density difference plots, Bader charge analysis, and DOS demonstrated the transfer of electrons from the \ce{Fe3O4} surface to murexide and was found to play a critical role in the adsorption process. This electron transfer pathway may have implications for the charge transfer mediation provided by murexide in \ce{Fe3O4} supercapacitor anodes at negative electrode potentials. Additionally, the band alignment analysis reveals that the lower energy levels of the LUMO and HOMO of the murexide molecule compared to the CBE and VBE of \ce{Fe3O4}, respectively, allows for the transfer of electrons from the \ce{Fe3O4} surface to murexide during adsorption, and may illustrate a conductive pathway for electrons to decrease electrode resistance.

Overall, these findings contribute to a deeper understanding of the interaction between murexide and \ce{Fe3O4} and have implications for the development of advanced energy storage systems. The demonstrated improvements in electrode performance and the insights into the adsorption process and charge transfer mechanisms provide a foundation for further exploration of murexide and its potential applications in various fields, including energy storage devices.

\begin{acknowledgement}
This research was funded by the Natural Sciences and Engineering Research Council of Canada, grant number RGPIN-2018-04014, and Faculty of Engineering of McMaster University. Calculations were performed using the Compute Canada infrastructure supported by the Canada Foundation for Innovation under John R. Evans Leaders Fund.
\end{acknowledgement}


\section*{Data availability}

The raw data (VASP input and structure files) required to reproduce computational findings are available in the Zenodo file repository \cite{Zenodo_10.5281/zenodo.8183854}.

\bibliography{literature}

\end{document}